\pdfoutput=0

\documentclass[sigconf,nonacm]{acmart}

\usepackage{algorithm}
\usepackage{alltt}
\usepackage{subfigure}

\AtBeginDocument{%
  \providecommand\BibTeX{{%
    \normalfont B\kern-0.5em{\scshape i\kern-0.25em b}\kern-0.8em\TeX}}}

\setcopyright{acmcopyright}
\copyrightyear{2021}
\acmYear{2018}
\acmDOI{10.1145/1122445.1122456}
\acmISBN{}
\acmConference[HPC Asia 2022]
{The International Conference on High Performance Computing in Asia-Pacific Region (HPC Asia 2022)}
{January 12--14 2022}{Online}
\acmBooktitle{HPCAsia2022: Proceedings of the International Conference on High Performance Computing in Asia-Pacific Region}

\begin{document}

\title{Strong Scaling of OpenACC enabled Nek5000 on several GPU based HPC systems}

\author{Jonathan Vincent}
\affiliation{%
  \institution{KTH Royal Institute of Technology}
  \streetaddress{100 44}
  \city{Stockholm}
  \country{Sweden}}
\email{jonvin@kth.se}

\author{Jing Gong}
\affiliation{%
  \institution{Uppsala University}
  \streetaddress{752 37}
  \city{Uppsala}
  \country{Sweden}}
\email{jing.gong@it.uu.se}
  
\author{Martin Karp}
\affiliation{%
  \institution{KTH Royal Institute of Technology}
  \streetaddress{100 44}
  \city{Stockholm}
  \country{Sweden}}
\email{makarp@kth.se}

\author{Adam Peplinski}
\affiliation{%
  \institution{KTH Royal Institute of Technology}
  \streetaddress{100 44}
  \city{Stockholm}
  \country{Sweden}}
\email{adam@mech.kth.se}

\author{Niclas Jansson}
\affiliation{%
  \institution{KTH Royal Institute of Technology}
  \streetaddress{100 44}
  \city{Stockholm}
  \country{Sweden}}
\email{njansson@csc.kth.se}

\author{Artur Podobas}
\affiliation{%
  \institution{KTH Royal Institute of Technology}
  \streetaddress{100 44}
  \city{Stockholm}
  \country{Sweden}}
\email{podobas@kth.se}

\author{Andreas Jocksch}
\affiliation{%
  \institution{CSCS - Swiss National Supercomputing Centre}
  \streetaddress{Via Trevano 131, 6900}
  \city{Lugano}
  \country{Switzerland}}
\email{andreas.jocksch@cscs.ch}

\author{Jie Yao}
\affiliation{%
  \institution{Texas Tech University}
  \streetaddress{2500 Broadway, TX 79409}
  \city{Lubbock}
  \country{USA}}
\email{jie.yao@ttu.edu}

\author{Fazle Hussain}
\affiliation{%
  \institution{Texas Tech University}
  \streetaddress{2500 Broadway, TX 79409}
  \city{Lubbock}
  \country{USA}}
\email{fazle.hussain@ttu.edu}

\author{Stefano Markidis}
\affiliation{%
  \institution{KTH Royal Institute of Technology}
  \streetaddress{100 44}
  \city{Stockholm}
  \country{Sweden}}
\email{markidis@kth.se} 

\author{Matts Karlsson}
\affiliation{%
  \institution{Link\"oping University}
  \streetaddress{581 83}
  \city{Link\"oping}
  \country{Sweden}}
\email{matts.karlsson@liu.se} 

\author{Dirk Pleiter}
\affiliation{%
  \institution{KTH Royal Institute of Technology}
  \streetaddress{100 44}
  \city{Stockholm}
  \country{Sweden}}
\email{pleiter@kth.se} 

\author{Erwin Laure}
\affiliation{%
  \institution{Max Planck Computing and Data Facility}
  \streetaddress{Gießenbachstraße 2, 85748}
  \city{Garching}
  \country{Germany}}
\email{erwin.laure@mpcdf.mpg.de}

\author{Philipp Schlatter}
\affiliation{%
  \institution{KTH Royal Institute of Technology}
  \streetaddress{100 44}
  \city{Stockholm}
  \country{Sweden}}
\email{pschlatt@mech.kth.se}


\renewcommand{\shortauthors}{Vincent, et al.}

\begin{abstract}
We present new results on the strong parallel scaling for the OpenACC-accelerated  implementation of the high-order spectral element fluid dynamics solver Nek5000. The test case considered consists of a direct numerical simulation of fully-developed turbulent flow in a straight pipe, at two different Reynolds numbers $Re_\tau=360$ and $Re_\tau=550$, based on friction velocity and pipe radius. The strong scaling is tested on several GPU-enabled HPC systems, including the Swiss Piz Daint system, TACC's Longhorn, J\"ulich's JUWELS Booster, and Berzelius in Sweden.  The performance results show that speed-up between $3$-$5$ can be achieved using the GPU accelerated version  compared with the CPU version on these different systems. The run-time for 20 timesteps reduces from $43.5$ to $13.2$ seconds with increasing the number of GPUs from $64$ to $512$ for $Re_\tau=550$ case on JUWELS Booster system. This illustrates the GPU accelerated version the potential for high throughput. At the same time, the strong scaling limit is significantly larger for GPUs, at about $2000-5000$ elements per rank; compared to about $50-100$ for a CPU-rank. 
\end{abstract}





\maketitle

\section{Introduction}
Among the many High-Performance Computing (HPC) frameworks that target high-fidelity Computational Fluid Dynamics (CFD) of incompressible flows, perhaps none is as salient as Nek5000~\cite{fischer2008nek5000}. Developed at MIT, Brown University and Argonne National Laboratory since the 1980s, Nek5000 is based on the high-order Spectral Element Method (SEM)~\cite{patera1984spectral} and is used in several important application domains, including (but not limited to) the study of turbulence in complex geometries such as airplane wings~\cite{vinuesa2018turbulent},  thermal-hydraulics in nuclear reaction cores~\cite{merzari2017large}, and ocean currents~\cite{ozgokmen2006product}. Several toolboxes around Nek5000 are available for specific tasks such as hydrodynamic stability analysis, turbulence statistics, uncertainty quantification etc. The Nek5000 package is today considered among the most used and important CFD frameworks in HPC for academic use.

Since its inception, Nek5000 has been written in Fortran 77 and has long relied entirely on the Message-Passing Interface (MPI)~\cite{walker1996mpi} for exploiting (or exposing) both inter- and intra-node parallelism. This strategy has historically worked remarkably well, particularly for HPC systems composed of homogeneous general-purpose processors (CPUs), where Nek5000 has demonstrated scalability to several thousands of nodes~\cite{offermans2016strong} and even won the prestigious Gordon Bell prize in 1999~\cite{tufo1999terascale}. 

Unfortunately, Nek5000 (in its current form) is ill-suited to exploit the increase in diversity and heterogeneity that the current HPC technology roadmap is embracing. With the end of Dennard's scaling (power scaling)~\cite{dennard1974design} and the impending termination of Moore's law (transistor scaling)~\cite{theis2017end}, future (Exascale~\cite{dongarra2011international}) architectures are expected to be more diverse and specialized than existing homogeneous systems we -- and, more importantly, Nek5000 -- have grown accustomed too. Today, seven of the ten systems in the world~\footnote{June 2021 Top500 list https://www.top500.org/lists/top500/2021/06/} are leveraging accelerator technology to provide the bulk of the compute capabilities. In particular, the highly-parallel Graphics Processing Units (GPUs) are commonly used, but there are ample reasons to believe that the future might be even more diverse (e.g., CGRAs~\cite{podobas2020survey}).

Today, there are many alternatives that could be considered to exploit accelerators from applications such as Nek5000. For example, prior work~\cite{karp2021High,karp2020optimization} have used both OpenCL~\cite{munshi2009opencl} and CUDA~\cite{Nvidia2007compute} for accelerating performance-critical (subject to Amdahl's law~\cite{amdahl1967validity}) kernels for both GPUs and Field-Programmable Gate Arrays (FPGAs), but this requires possible a lot of rewriting (OpenCL) or yields less portable solutions (CUDA). Increased portability can be gained by Just-in-Time (JIT) compiling performance-critical kernels onto accelerators~\cite{fischer2021nekrs} but is potentially vulnerable to low performance subject to the maturity of the JIT compiler. A third option, which is also the option we pursue in this paper, is to use OpenACC directives and API for CUDA Fortran kernels. The aspiration of OpenACC is to maintain portability (with varying levels of success~\cite{deakin2019performance}) through the use of directives and let the compiler transform said directives into high-performance accelerator (or general-purpose) code. An initial port of Nek5000 to OpenACC, however without CUDA kernels, has been described by Otero et al. \cite{otero_gong_min_fischer_schlatter_laure_2019}. The present paper constitutes a major improvement over those results, including scaling results on large-scale architectures and realistic flow cases.

In this paper, we describe our efforts in modernizing Nek5000 using the OpenACC programming model. In summary, our paper provides the following contributions:
\begin{enumerate}
\item We developed and validated an OpenACC-version of Nek5000 and describe the implementation and techniques used within,
\item We empirically quantify the scalability of our OpenACC-Nek5000 implementation compared to the existing state-of-the-art, and
\item We analyze the performance of our Nek5000 implementation and identify future performance opportunities and limitations.
\end{enumerate}

In the remainder of the paper, we first describe the
theoretical background of the Nek5000 code, then we present the
results of Nek5000 running on a variety of GPU based systems running a case of a turbulent flow in a straight pipe. Finally we
present the conclusions and future work.

\section{Mathematical background}

Nek5000 integrates in time the incompressible Navier--Stokes equations consisting of the momentum and continuity equations:

\begin{equation}
  \frac{\partial \mathbf{u} }{\partial t} +
  \left(  \mathbf{u} \cdot \nabla \right) u =
  - \nabla p + \frac {1}{Re} \nabla^2 \mathbf{u} +
  \mathbf{f} \ ,
\end{equation}

\begin{equation}
\nabla \cdot  \mathbf{u} = 0 \ ,
\end{equation}
where $\mathbf{u}$ is the velocity, $p$ the pressure (divided by density), $Re$ the Reynolds number and $\mathbf{f}$ is a forcing term. The Reynolds
number $Re_\tau = \frac{UL}{\nu}$ is a function of a typical velocity scale
$U$, length scale $L$ and kinematic viscosity $\nu$. There are multiple ways this problem can be discretized for numerical modelling, however one  crucial aspect is the proper treatment of the spurious pressure modes. To deal with this Nek5000 supports two different formulations relying on staggered ($\mathbb{P}_N - \mathbb{P}_{N -2}$) and collocated (
$\mathbb{P}_N - \mathbb{P}_{N}$) grids.  In this paper we consider the more traditional $\mathbb{P}_N - \mathbb{P}_{N -2 }$ formulation  which is similar to a classical velocity-correction method.

The momentum equation is time integrated via an implicit--explicit
scheme, also known as BDF$k$-EXT$k$ (Backward Difference Formula and
Extrapolation of order $k$). This can be illustrated in a semi-discrete way as
\begin{equation} \label{Fn_def}
\sum^k_{j=0} \frac{b_j}{\Delta t} \mathbf{u}^{n-j} =
- \nabla p^n  + \frac{1}{Re} \nabla^2 \mathbf{u}^n
+ \underbrace{ \sum^k_{j=1} a_j
  N \left( \mathbf{u}^{n-j} \right)
  + \mathbf{f}^n }_{
  \mathbf{F}_n 
    \left( \mathbf{u} ,  \mathbf{f}\right)},
\end{equation}
where we denoted the nonlinear operator $\mathbf{u} \cdot \nabla
\mathbf{u} = N(\mathbf{u} ) $ and $b_k$ and $a_k$ are the coefficients
of the implicit time derivative discretization and explicit
extrapolation, respectively. Following the pressure correction method of Perot~\cite{Perot1993} this equation is solved in three steps: the solution of the Helmholtz problem for the intermediate velocity $\mathbf{u}^{*}$
\begin{equation}
\left( \frac{b_0}{\Delta t} - \frac{1}{Re} \nabla^2 \right) \mathbf{u}^{*} = \mathbf{F}_n \left( \mathbf{u} ,  \mathbf{f}\right) - \sum^k_{j=1} \frac{b_j}{\Delta t} \mathbf{u}^{n-j} - \nabla p^{n-1},
\end{equation}
followed by the calculation of the pressure update $\delta p = p^n -p^{n-1}$
\begin{equation}
\frac{\Delta t}{b_0} \nabla^2 \delta p = - \nabla \mathbf{u}^{*},
\label{poisson}
\end{equation}
and concluded with the actual correction of both final pressure and velocity
\begin{eqnarray*}
    p^n &=& p^{n-1} + \delta p, \\
    \mathbf{u}^{n} &=& \mathbf{u}^{*} + \nabla \delta p.
\end{eqnarray*}
The two first steps are performed using iterative solvers with proper preconditioners and acceleration techniques such as the projection of a current solution at each time step onto a subspace spanned by previous solutions~\cite{Fischer1998}. 

The most expensive step is -- as for all incompressible flow solvers -- the solution of the (consistent) Poisson equation for the pressure Eq.~\ref{poisson}, which is the main source of stiffness. This problem requires a specific preconditioner based on the additive overlapping Schwarz method given by
\begin{equation}\label{schwarz}
    M_0 ^{-1} := R_0^T A_0^{-1} R_0 + \sum _{k=1} ^K 
    R_k ^T  A_k  ^{-1} R_k \ .
\end{equation}

Here solutions of the local Poisson problems in overlapping subdomains $R_k^T  A_k^{-1} R_k$ are combined with the coarse grid problem $R_0^T  A_0^{-1} R_0$, which is solved on few degrees of freedom, but covers the entire domain. The overlapping subdomain calculation is naturally parallizable. However the coarse grid solve is much more difficult to do in parallel, and so has not been moved from the CPU onto the GPU. 

In this paper we will consider what is known as the  XXT method for solving the coarse grid solve. In this method a Cholesky factorization of the matrix $A_0^{-1}$ into the form $X X^T$, with a convenient refactoring of the underlying matrix to maximise the sparsity pattern of $X^T$~\cite{Tufo2001} is used.

The spatial discretization in Nek5000 is based on the spectral element method~\cite{patera1984spectral}. In the spectral element method we decompose the computational domain into a set of non-overlapping, body-conforming hexahedral subdomains called elements. Each element is treated as a spectral domain with  finite-dimensional sub-spaces spanned by a tensor product of the one-dimensional Lagrange interpolation polynomials. For staggered meshes  the Lagrange interpolants of order $N$ (applied to the Gauss-Lobatto-Legendre points) and $N-2$ (applied to the Gauss-Legendre points) are used for velocity and pressure respectively. An important aspect is the natural element-wise domain decomposition, and use of the tensor products, which allows to separate global (exchange of element face values) and local (vector-matrix multiplication) operations and to perform the local ones efficiently in the matrix-free form.

\section{GPU Implementation and Optimization}

Current architectures show a trend towards more and more parallel execution units on the node, currently a few thousand CUDA cores per Nvidia GPU. Naturally, for optimal performance, applications should make good use of these available cores. The many cores push the development of highly parallel algorithms, but the increased amount of intra-node parallelism also requires special features of the hardware compared to conventional CPUs. On GPUs, this has manifested in several ways. In particular, core components of GPUs are very fast local memories called shared memory, which is basically a programmable cache, and fast atomic instructions. Atomic operations for floating-point were originally supported for single precision only and in the past, significant considerations were necessary to overcome the absence of such features in double-precision \cite{hariri2016portable}. Later generation cards have begun to support double-precision arithmetic natively and an incredible throughput can be achieved. This makes modern GPUs appealing for CFD applications as most current CFD codes such as Nek5000 are entirely based on double-precision computations, even though there are studies relaxing this requirement at least partially.

In our GPU implementation we map the numerical algorithms of the fluid dynamics application to the hardware. We assume an ideal programming model, where the algorithms are mapped to the hardware without performance loss due to software layers. We determine how much shared memory per streaming multiprocessor (SMX) is needed for which parameters of the algorithm. Furthermore we design a performance model in order to determine the number of GPUs to be placed on one node for a given network for typical productions runs.

The GPU implementation was created from the CPU implementation using OpenACC, which was relatively straightforward addition of directives to the existing CPU version. There were several complications however. The small maths kernels are often called repeatedly in loops, so significant gains can be made by moving the loop inside the kernel, at some cost to code complexity as the generic maths function has become more specific with the loop.  We also found that calculations with reductions can be quite inefficient when implemented in OpenACC, so replacing those directives with hand written kernels can improve performance. We also found that OpenACC can be very conservative with data movement, so there can be considerable scope to reduce the data copying between CPU memory and GPU memory by careful use of data regions and moving some additional calculations to the GPU.

\subsection{Communication cost}
\begin{figure}
\begin{center}
\includegraphics[width=0.45 \textwidth]{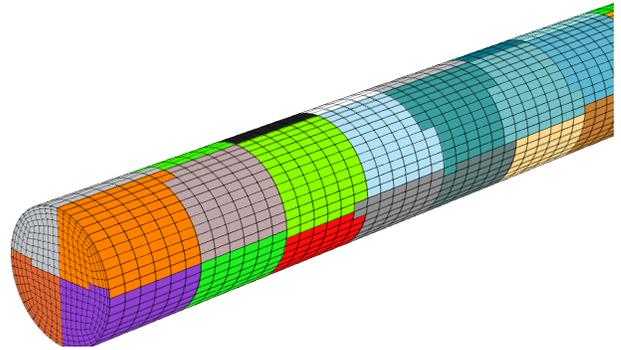}
\caption{Partitioning of the elements on 64 processes in the pipe simulation with $Re_{\tau} = 180$ (reproduced from \cite{offermans2016strong})}
\label{fig:pipe_partitioning}
\end{center}
\end{figure}

One factor of the communication cost is the geometry and topology of the problem at hand. We focus here on the flow in a straight periodic pipe and compare it with a cube with periodic boundary conditions in all directions. The strong scaling properties of Nek5000 have been investigated already in general and for the pipe flow specifically \cite{fischer2015scaling,offermans2016strong}. Strong scaling is not our only one focus since GPU equipped computers are designed for high throughput. The pipe and the reference cube are decomposed with the recursive spectral graph partitioner typically used for Nek5000 due to its optimal element distribution. Furthermore we compare with a Fourier mode discretisation of the cube. In this case we assume a slab or pencil decomposition of the computational domain with a transposition done by an all-to-all collective communication \cite{jocksch2019optimized}. Figure~\ref{fig:pipe_partitioning} shows the results of the partitioning for the pipe and the cube. The pipe has less edge cuts and communication volumes than the cube for the same number of partitions. This is mostly attributed to the non-periodic boundary conditions but also to the different shape of the computational domain. Therefore we can consider the cube as a conservative estimate for the modelling of communication cost.

\subsection{Network and node size}
Recently, multiple GPUs within a compute node can be connected directly using technology such as NVLINK~\cite{foley2017ultra}, thus bypassing (in part) the need to communicate through (expensive) PCIe.  Such functionality and flexibility, in turn, raise the following question: what is a good machine balance in terms of both the node-level CPUs-to-GPUs ratio and network from the perspective of our Nek5000 application? For this optimisation problem we would like to give a hint from the prospective of our application. In advance we will discuss two simplified cases for domain decomposition and communication patterns. For a 1d decomposed mesh with a block size matching on one GPU one could merge neighbouring blocks to a blocksize of $n$ GPUs in order to account for $n$ times larger nodes. As a consequence the traffic between neighbouring domains would remain unchanged and the overall traffic would decrease by a factor of $n$. For a 3d decomposition of the mesh the situation is a different one. We imagine every block to be a cube and merge 8 of those cubes into one of double edge length. The message size to be communicated between neighbours is four times as big for the large cube compared to the small one. However, the overall traffic reduces by a factor of two.

Our real case is more complex, the mesh is decomposed using a graph partitioner. We model the communication volume for small and large nodes dividing the mesh into small and large subdomains, respectively. During a single run different virtual node sizes are considered by counters incremented depending on whether messages are passed within a virtual node or between virtual nodes.

\subsection{CUDA Fortran kernels}

Efficient utilization of the cache and shared memory on GPUs are essential for high performance. Even if the spectral element method has a high operational intensity compared to other methods, the core computations are still in the memory-bound domain. Shared memory kernels were investigated for Nek5000 already in \cite{gong2016nekbone}, but we focus on a further developed solution. At the level of a single GPU, the high performance for the nested loops (e.g. those in Alg. \ref{alg:axhelm}) are crucial. The memory access pattern for the single spectral element is irregular, but data is reused which means that shared memory can be utilized efficiently. Also, we observe that storing temporary arrays in shared memory rather than global memory yields a significant performance improvement. As these arrays are temporary, storing them in global memory is not necessary, and execution time can be improved. With this in mind, we observe that the maximum size of the spectral elements if we want to obtain high performance, is directly linked to the capacity of the shared memory. The hardware used -- an Nvidia V100~\cite{choquette2018volta} GPU card  -- has the property that an overall size of 128kB of fast memory can be set up to a part as shared memory, and the rest is L1 cache. In order to estimate the ideal size of such fast memory, we reduce it artificially by declaring a shared memory array that is not used.

\begin{figure}
\begin{center}
\includegraphics[width=0.47\textwidth]{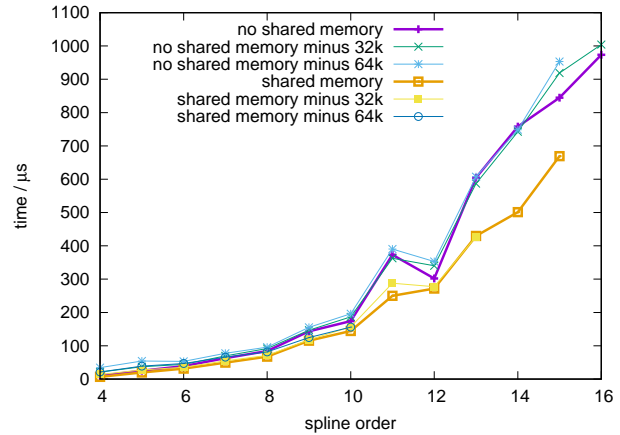}
\caption{Execution time for 1000 spectral elements as a function of element polynomial order $N$  using shared memory and no shared memory}
\label{fig:shared_nonshared}
\end{center}
\end{figure}

Fig.~\ref{fig:shared_nonshared} shows the execution time of the kernel for an implementation using shared memory and not using shared memory.
The shared memory option is always the faster one. However the polynomial order is restricted to $14$. Larger sizes of the elements can be accommodated, but only without using shared memory.

To exploit the shared memory on GPU for the lower polynomial orders which are applied to real physical simulation problems, we have implemented several CUDA Fortran kernels for the most time consumption subroutines, which are called using the OpenACC clause. Table \ref{alg:axhelm} shows the CUDA Fortran kernel for subroutine \texttt{axhelm}. The subroutine requires matrix-vector productions to apply for the discrete Hmholtz operator. For the case of $7^{th}$ polynomial order with $lx=8$, two 2D-arrays and three 3D-arrays with double precision take around $13$KB RAM that satisfies most modern GPUs.    

\begin{algorithm}[!htp]
 \caption{CUDA Fortran kernel for subroutine \texttt{axhelm}. \label{alg:axhelm}}
 \begin{alltt}
 ...
      real, shared :: shdxm1(lx1,ly1)
      real, shared :: shdxtm1(lx1,ly1)
      real, shared :: shur(lx1,ly1,lz1)
      real, shared :: shus(lx1,ly1,lz1)
      real, shared :: shut(lx1,ly1,lz1)
      
      rtmp = 0.0
      stmp = 0.0
      ttmp = 0.0
      do l = 1, lx1
         rtmp = rtmp + shdxm1(i,l)  * u(l,j,k,e)
         stmp = stmp + shdxm1(j,l)  * u(i,l,k,e)
         ttmp = ttmp + shdxm1(k,l) * u(i,j,l,e)
      enddo
      wr = g1xyz(i,j,k,e)*rtmp + g4xyz(i,j,k,e)*stmp
     $       + g5xyz(i,j,k,e)*ttmp
      ws = g2xyz(i,j,k,e)*stmp + g4xyz(i,j,k,e)*rtmp
     $       + g6xyz(i,j,k,e)*ttmp
      wt = g3xyz(i,j,k,e)*ttmp + g5xyz(i,j,k,e)*rtmp
     $       + g6xyz(i,j,k,e)*stmp
      htmp = h1(i,j,k,e)
      shur(i,j,k) = wr * htmp
      shus(i,j,k) = ws * htmp
      shut(i,j,k) = wt * htmp

      call syncthreads()

      wijke = 0.0
      do l = 1, lx1
         wijke = wijke + shdxtm1(i,l) * shur(l,j,k)
     $                 + shdxtm1(j,l) * shus(i,l,k)
     $                 + shdxtm1(k,l) * shut(i,j,l)
      enddo
      w(i,j,k,e) = wijke
...
 \end{alltt}
 \vspace*{-0.5cm}
\end{algorithm}

\begin{figure}[htp!]
  \centering
   \subfigure[]{
    \includegraphics[width=0.47\textwidth]{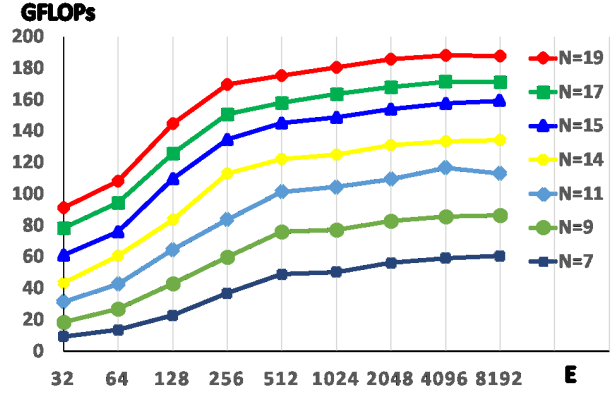} \label{fig:nekbone_v100}
   }
     \subfigure[]{
       \includegraphics[width=0.47\textwidth]{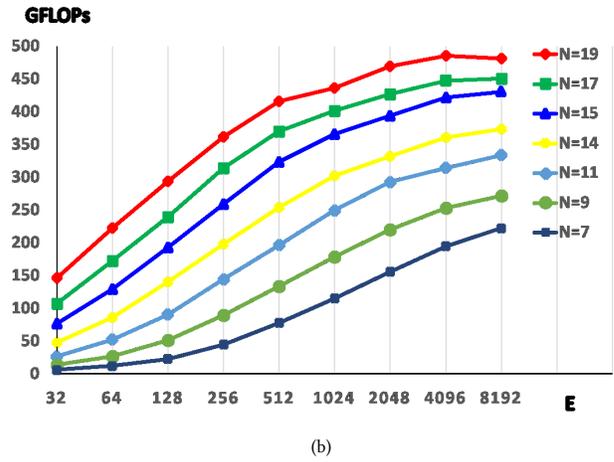} \label{fig:nekbone_p100}
       }
  \caption{The performance result of Nekbone on a single P100 \subref{fig:nekbone_v100} and V100 \subref{fig:nekbone_v100} GPU. $E=32-8192$ is the number of elements while $N=7-19$ is the polynomial order}
  \end{figure}

With this optimization, maximum performance of $480$ GFlops for the Nek5000 mini-app, Nekbone, can be achieved. That is almost double that on a Nvidia P100 GPU,  \ref{fig:nekbone_p100} and~\ref{fig:nekbone_v100}. The performance of Nek5000 mini-app, Nekbone much depends on the number of elements and polynomial order, which is as expected. 

\section{Experimental Setup}
We executed experiments on turbulent flow in a pipe on four different machines in order to assess the scaling and performance of our OpenACC version of Nek5000. 
\subsection{Hardware}
The machines tested were Piz Daint at CSCS, Longhorn at TACC,  Berzelius at NSC, and the JUWELS Booster system at Juelich. The details of each machine are as follows. First,  Piz Daint is a system with 5704 Cray XC50 compute nodes in the GPU partion. Each XC50 node has one 12 core Intel Xeon E5-2690 v3 CPU clocked at 2.60GHz and a Nvidia Tesla P100 GPU with 16GB of HBM memory with up to 720GB/s memory bandwidth. As for the interconnect, the nodes are connected using the Cray Aries network with a dragonfly topology. Second, Longhorn is an IBM system equipped with two 20 core IBM Power 9 processors per node. In addition, each node has four Nvidia V100 GPU with 16GB of HBM memory and a maxmimum memory bandwidth of 900GB/s. The nodes are connected using Mellanox EDR Infiniband using a spine-and-leaf topology. Third, we used Berzelius, which is a Nvidia SuperPOD system where each node has two AMD Epyc 7742 CPUs with 64 cores per processor and eight Nvidia Tesla A100 GPUs per node. The A100 GPUs each have 40 GB of RAM and a peak bandwidth of 1550GB/s. The nodes are connected using eight Mellenox HDR cards Infiniband, using a fat tree topology. Lastly, we used the JUWELS Booster system is a Sequana system from Atos where each node has two 24 core AMD EPYC 7402 processors and four Nvidia A100 GPUs. The A100 GPUs are equipped with the same memory system as in Berzelius. The nodes in the JUWELS Booster system are connected using four Mellenox HDR cards per node and using a Dragonfly+ topology~\cite{shpiner2017dragonfly}. For our CPU runs we also used the JUWELS Booster System. For all experiments we used one MPI rank per GPU and all the GPUs on each allocated node.

\subsection{Flow case}
\begin{figure}[htp!]
    \centering
    \subfigure[]{
       \includegraphics[width=0.10\linewidth]{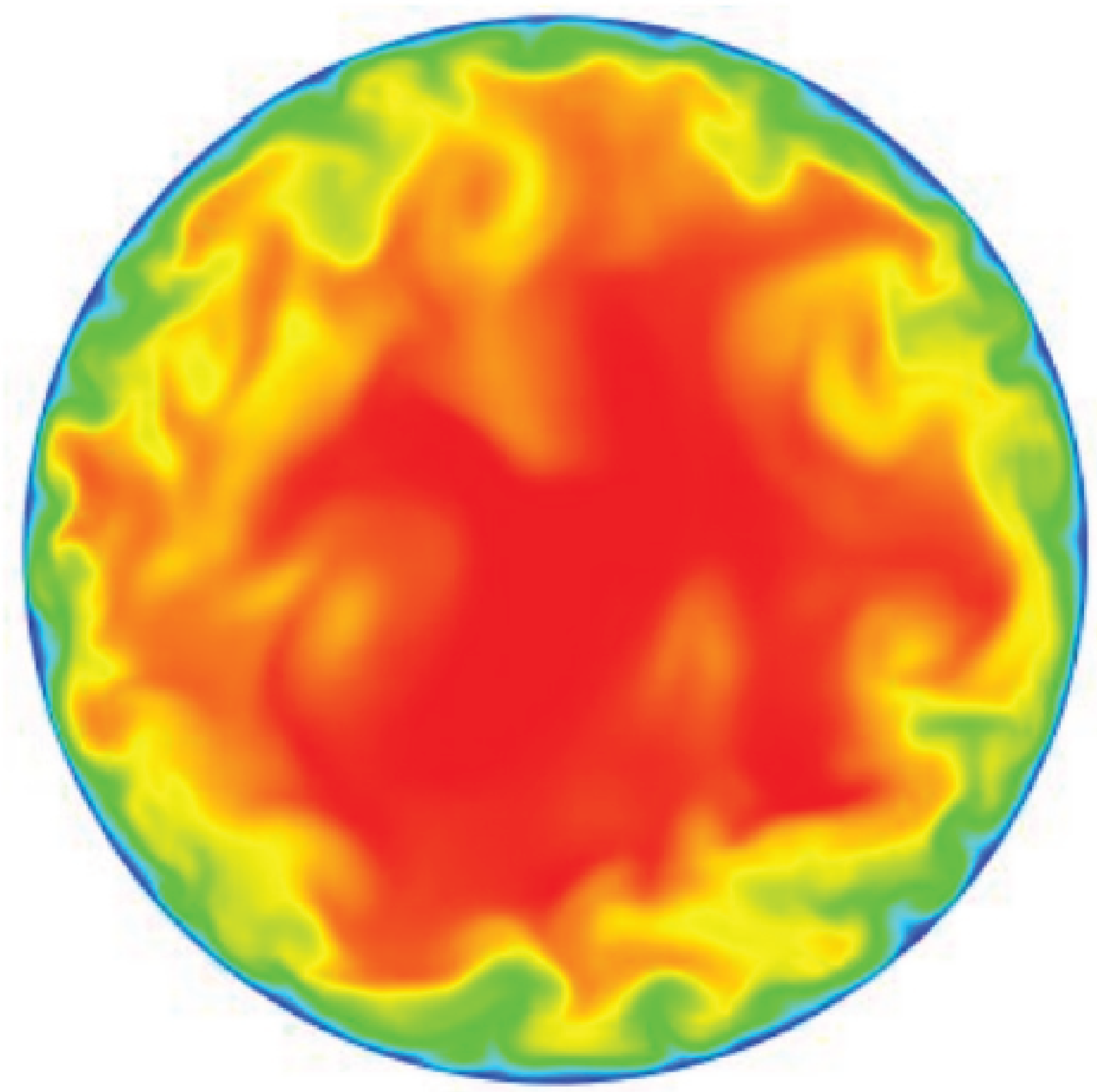}
       }
    \subfigure[]{
     \includegraphics[width=0.8\linewidth]{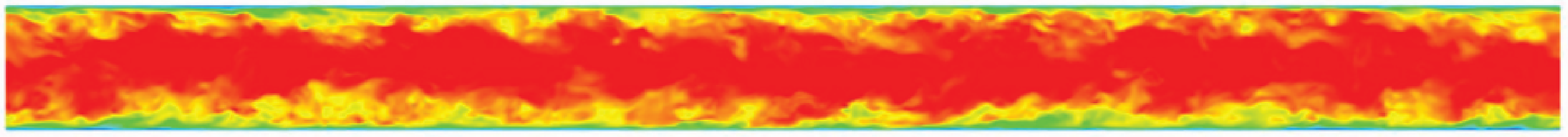}
     }
 \caption{Visualization of a pipe simulation with $Re_\tau=360$ (reproduced from \cite{schlatter2012PDC}). Shown are colors of the streamwise velocity component, ranging from fast (red) to zero (green). (a) cross-sectional view, (b) vertical cut through the pipe axis. The figure highlights the small details characteristic of turbulent flow, which ultimately lead to the stringent resolution requirements.} \label{fig:pipe}
\end{figure}

The flow case we consider is the fully-developed turbulent flow in a straight pipe. A thorough description of the flow  configuration as well as a detailed analysis of the physical results can be found in \cite{elkhoury_schlatter_noorani_fischer_brethouwer_johansson_2013,offermans2016strong}, see Figure \ref{fig:pipe}, with results at four different friction Reynolds numbers $Re_{\tau} = 180$, $360$, $550$ and $1000$. The friction Reynolds number, also known as the K\'arm\'an number, is defined as $Re_{\tau} = u_{\tau} R / \nu$, where $u_{\tau}$ is the friction velocity, $R$ is the radius of the pipe, and  $\nu$ is the kinematic viscosity. The bulk Reynolds number is defined as $Re_{b} = 2 U_b R / \nu$, where $U_b$ is the mean bulk velocity.  A summary of the different simulations and associated number of elements and number of grid points is presented in \cite{offermans2016strong}. The specific cases relevant for this study  were run at  Reynolds numbers $Re_\tau = 360$ and required $237120$ total elements (shown in Fig.~\ref{fig:pipe}) and $Re_\tau=550$, which corresponds to $853632$ total elements. It is interesting to note that these specific Reynolds numbers correspond to the speeds and dimensions comparable to household plumbing (e.g.\ water faucets).

\begin{figure}[htp!]
  \centering
   \subfigure[]{
    \includegraphics[width=0.47\textwidth]{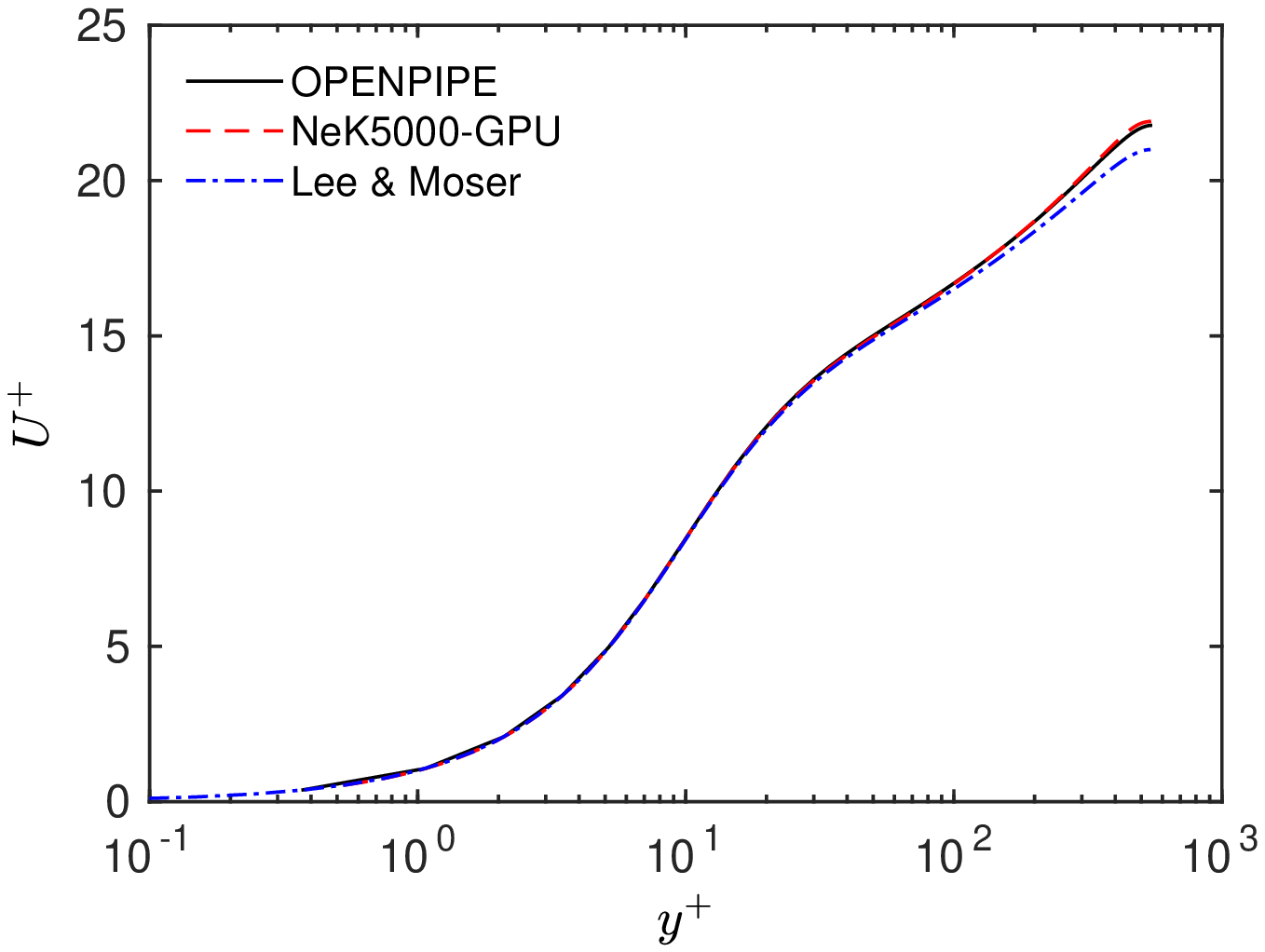} \label{fig:um_re550}
   }
     \subfigure[]{
       \includegraphics[width=0.47\textwidth]{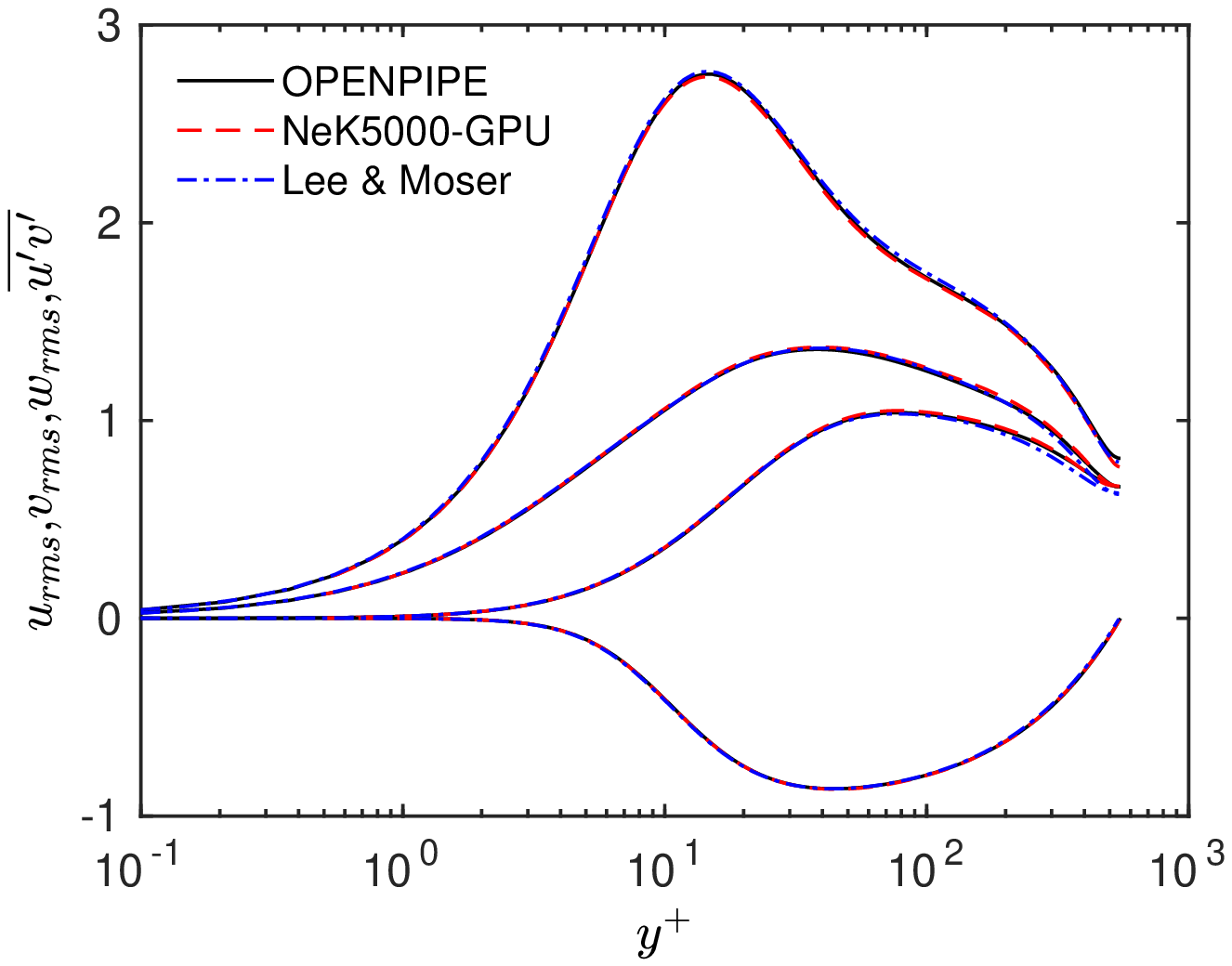} \label{fig:urms_re550}
       }
  \caption{Comparison of turbulence statistics for the pipe flow and channel flow using different  algorithms and codes. $Re_\tau=550$ for Nek5000 using $128$ P100 GPUs on Longhorn. (a) Mean flow profile, (b) turbulence fluctuations.} \label{fig:statistics}
\end{figure}

In order to validate this OpenACC version, we compare the results with the
DNS (direct numerical simulation) for wall-boundary turbulence performed by Lee and Moser (2015) \cite{myoungkyu2015JFM} for channel flows in larger domains of size $L_x\times L_z=8\pi h\times 3 \pi h$ using a B-Spline method, as well as the open source code \texttt{OpenPipeFlow} \cite{openpipeflow} using compact finite differences. For evaluating the turbulence statistics, we ported the statistics toolbox for turbulent pipe flow developed in \cite{saleh2019KTH} to GPU. Fig.~\ref{fig:statistics} presents the excellent agreement of the computed profiles of the first and second order statistics of velocities between pipes and channels. Further discussions on the expected agreement (and expected disagreement) can be found in El Khoury et al. \cite{elkhoury_schlatter_noorani_fischer_brethouwer_johansson_2013}.

\section{Performance results}
\begin{figure}
  \includegraphics[width=0.47\textwidth]{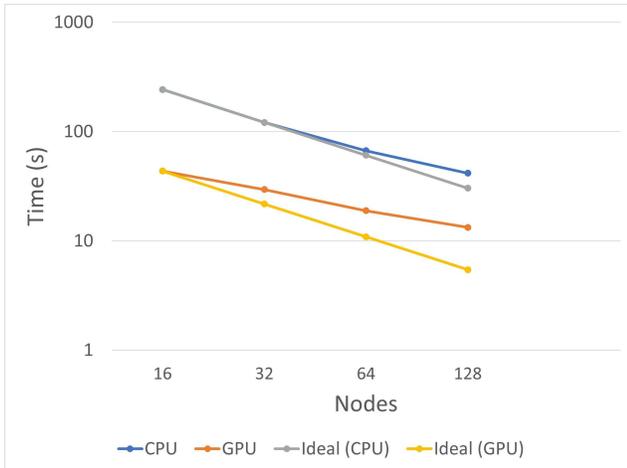}
\caption{\label{CPUvsGPU} CPU and GPU performance for $Re_\tau =550$ with maximum polynomial order $N=9$, on JUWELS Booster. Ideal scaling based on the  16 node results is also shown.}
\end{figure}

The run-time in seconds for all tests is measured between $30$ and $50$ timesteps. Fig.~\ref{CPUvsGPU} shows the comparison between the CPU performance and GPU performance per node on the JUWELS Booster system for the $Re_\tau=550$ case with polynomial order of 9 for the velocity. For the 128 nodes case, there are 139 elements per rank for the CPU case and 1668 elements per rank for the GPU case. As can be seen from the results, the OpenACC GPU code runs approximately five times faster for a smaller number of nodes, decreasing to a little over three times faster for the larger number of nodes (and GPUs) used. The decreased speed up as the number of GPUs increases is partly explained by increased communication overhead, but in particular by the decreased performance of the GPUs when the number of points per GPU decreases. This is made clear in Fig.~\ref{fig:nekbone_p100} and \ref{fig:nekbone_v100} where the performance quickly decreases with the number of elements. This behaviour is currently the main limiting factor for strong scaling on multiple GPUs, but as is shown, the absolute runtime is still improved over the CPUs. This  scaling behavior was also recently discussed by Fischer et al. for various different PDE solvers in \cite{fischer2020scalability}.

\begin{figure}[htp!]
  \centering
   \subfigure[]{
    \includegraphics[width=0.47\textwidth]{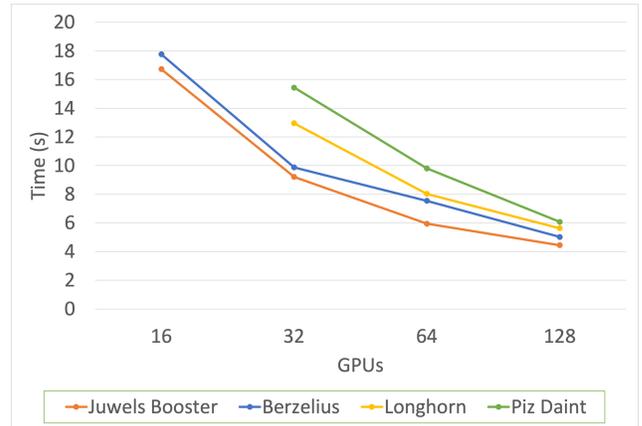} \label{Retau360_8}
   }
     \subfigure[]{
       \includegraphics[width=0.47\textwidth]{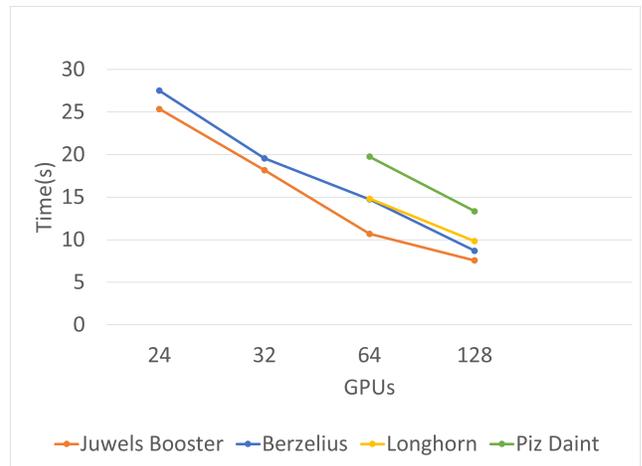} \label{Retau360_10}
       }
  \caption{Results for Reynolds number $Re_\tau=360$ and maximum polynomial order of  $N=7$ \subref{Retau360_8} and 9 \subref{Retau360_10}.}
  \end{figure}

Figs.~\ref{Retau360_8} and \ref{Retau360_10} show the results for $Re_\tau  = 360$ on the various systems with maximum polynomial order of 9 and 7, respectively. Due to the significantly less memory available per GPU on the P100 GPUs (16 GB) and the V100 GPUs (16 GB) than the A100 GPUs (40 GB), the minimum number of GPUs required for there to be enough GPU memory to be available is significantly higher on Piz Daint and Longhorn than for Berzelius and JUWELS Booster. Overall the performance improvement between Piz Daint and JUWELS
Booster is approximately a factor of two, which is consistent with the increase in the double precision flops and bandwidth from the 4.7
TFlops and 732 GB/s for the P100 GPUs and 9.7 TFlops and 1550 GB/s bandwidth of the A100 GPUs.

For the two A100 systems Berzelius and JUWELS Booster, we see that the
higher performance network JUWELS Booster system gives it superior performance. This advantage increases as the number of nodes
increases.

\begin{figure}[htp!]
  \centering
   \subfigure[]{
    \includegraphics[width=0.47\textwidth]{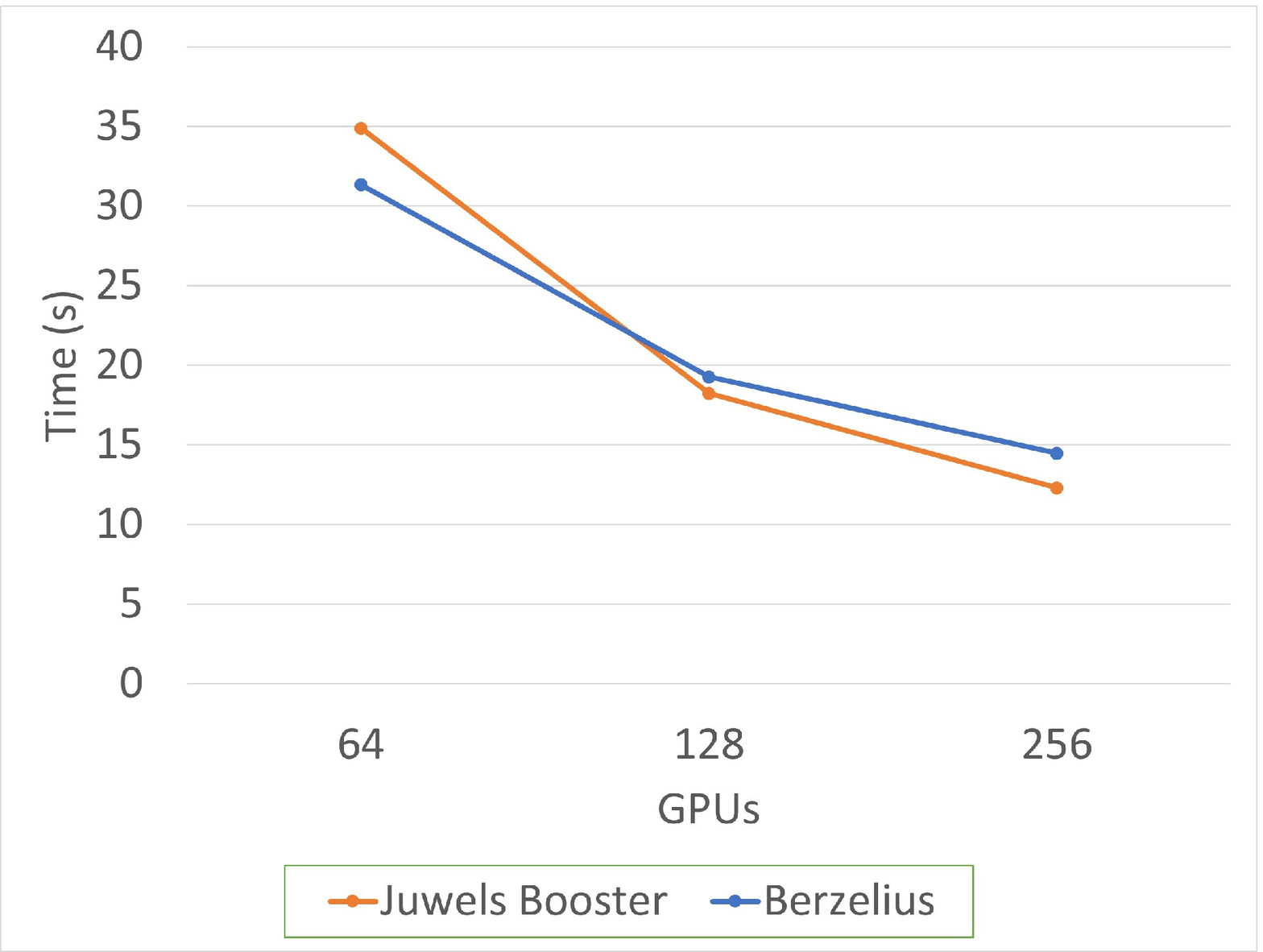} \label{Retau550_8}
   }
     \subfigure[]{
       \includegraphics[width=0.47\textwidth]{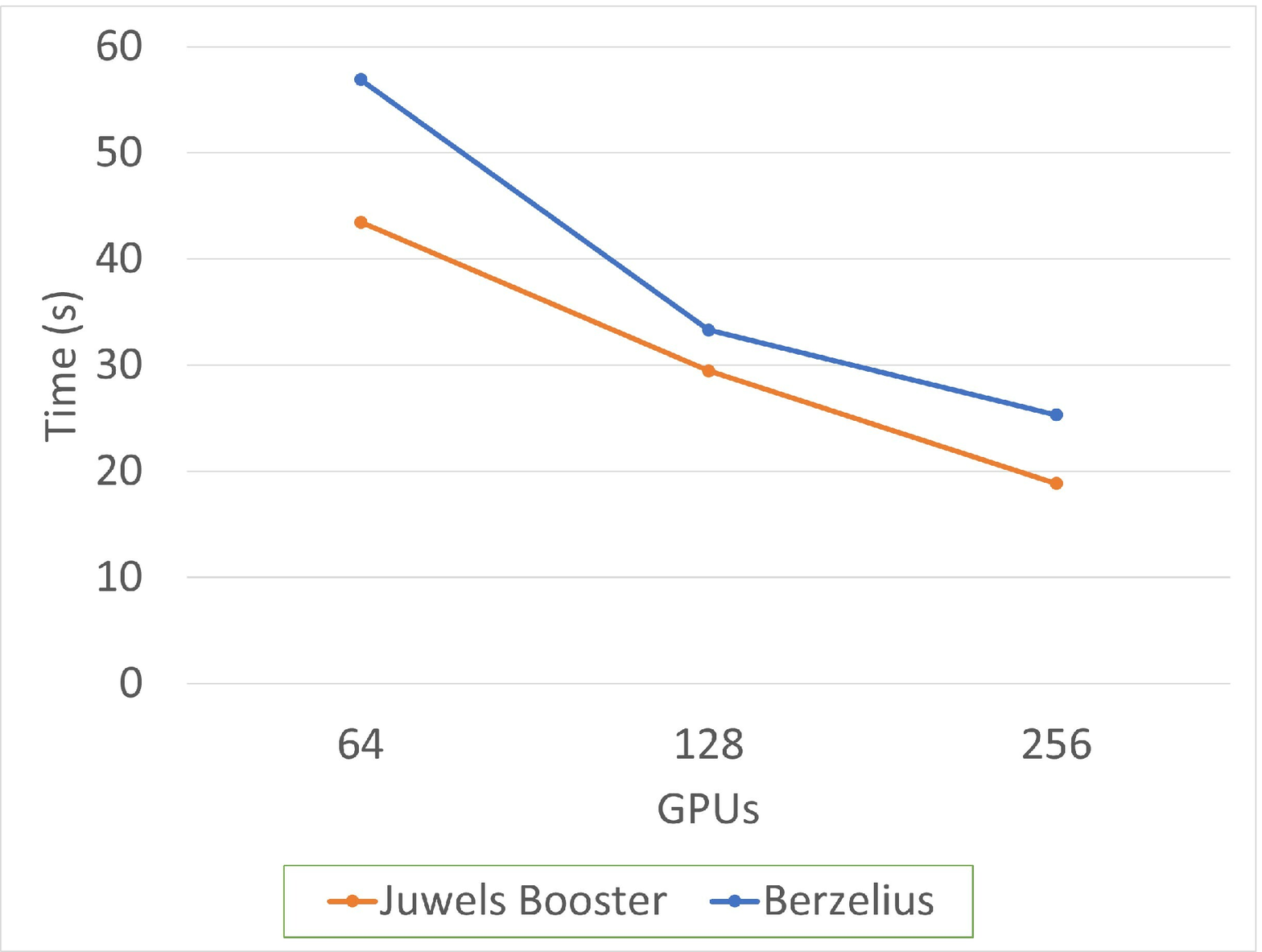} \label{Retau550_10}
       }
  \caption{GPU-based performance results for Reynolds number $Re_\tau=550$ and maximum polynomial order of  $N=7$ \subref{Retau550_8} and 9 \subref{Retau550_10}.}
  \end{figure}

Fig.~\ref{Retau550_8} shows the results for $Re_\tau  = 550$ and with a maximum polynomial order of 7. Similarly, Fig.~\ref{Retau550_10} presents the performance for a maximum polynomial order in the x,y and z directions as 9. Due to technical issues on Piz Daint and lower amount of memory of the V100 GPUs compared to the A100 GPUs, and limited number of nodes of the Longhorn system, simulations with $Re_\tau  = 550$ were not performed on these systems.

For the simulations with $Re_\tau  = 550$, we see a similar performance to the $Re_\tau  =360$ results, where the improved network performance of the JUWELS Booster system gives overall improved performance. The one exception is
$Re_\tau =550$ with maximum polynomial order of 9, where reduced number of nodes for the Berzelius system with 8 GPUs per node gives an advantage over the 4 GPUs per node of JUWELS Booster.

Overall the speedup on GPU is less than would be predicted from the increase in Flops from using GPUs instead of CPUs. This is mainly due to the remaining CPU only work in the coarse grid solve. The MPI performance is in general good, similarly to the CPU only version. Some more work is also needed to reduce CPU to GPU data copies.

\section{Related Work}
The acceleration of the spectral element method on GPUs has been pursued in multiple directions by several groups. The acceleration of Nek5000 with OpenACC directives was first explored by Markidis et al. by accelerating the mini-app Nekbone in \cite{markidis2015openacc} and then improved with CUDA Fortran implementations for the core computations by Gong et al. in \cite{gong2016nekbone}. In the most recent development of Nekbone on GPUs, the work of Świrydowicz et al. \cite{swirydowicz2019acceleration} was incorporated into the mini-app by Karp et al. \cite{karp2020optimization}. Acceleration of the core tensor operation in Nekbone on novel architectures such as FPGAs and a comparison between CPUs and GPUs was also extensively evaluated in \cite{karp2021High}. Finally, initial results for our OpenACC version of the entirety of Nek5000 were shown in \cite{otero_gong_min_fischer_schlatter_laure_2019}. In this article we showcase the performance and scaling of the entire Nek5000 solver on GPUs with OpenACC directives and CUDA Fortran implementations for performance-critical kernels. 

A complete rewrite of Nek5000 in C++, targeting multiple different backends has also been introduced by Fischer et al. \cite{fischer2021nekrs}, termed NekRS. Unlike our OpenACC version, this is a complete revision of the original Fortran code, relying heavily on Just in Time Compilation (JIT) with the help of OCCA \cite{medina2014occa}. 

In other works, different approaches to port similar codes to GPUs and other architectures have been explored. In Neko~\cite{jansson2021neko}, another CFD solver based on the spectral element method, the authors instead utilize a modular codebase in modern Fortran to accommodate different accelerators rather than a specific programming framework. More generally, Eichstädt et al. recently evaluated several different approaches for performance portability in the high order finite element solver Nektar++~\cite{eichstadt2020comparison}. 

\section{Conclusions and future work}

We have presented  various performance results for Nek5000 on GPUs.  Our GPU
version using OpenACC shows a significant performance improvement compared to the CPU version per node, where the performance per node improved by a factor of $3$ to $5$ with the larger improvement at low numbers of nodes. Note that these results are obtained for turbulent pipe flow, which constitutes a realistic large-scale flow case. We can therefore conclude that Nek5000 can successfully be used on modern GPU clusters.

The underlying spectral element discretization typically scales very well on parallel systems as there is significant work per element; therefore, most of the operations could be mapped with good efficiency to GPUs. There is one exception that could not be mapped to the GPU, though: The coarsest grid level of the pressure multigrid preconditioner, where the pressure is solved globally on a coarser grid. As we increased the number of GPUs used, we observed that this part of the algorithm begins to dominate the time step. The coarse grid solver operates on a low number of grid points per rank, and involves sparse Cholesky factorization. The lack of parallelism and strided memory accesses makes it a significant challenge to get good performance in parallel on both CPU and GPU systems. However, as discussed in the paper, larger element counts per node can alleviate this problem. Therefore, the strong scaling limit is significantly larger for GPUs, and can be estimated for JUWELS to be at about $2000-5000$ elements per rank; compared to about $50-100$ for a CPU-rank. Note that a rank corresponds to one GPU or a CPU core, respectively.

As the coarse grid is the current major bottleneck, we believe that future work should focus on an improved algebraic multigrid preconditioner with GPU implementations such as AmgX~\cite{AMGXweb} or HYPRE~\cite{HYPREweb}. With an improved coarse grid solver it is expected that additional performance gains at lower element counts are possible.

\begin{acks}
Financial support was provided by the SeRC Exascale Simulation Software Initiative (SESSI), the European Commission Horizon 2020 project grant ``EXCELLERAT: The European Centre of Excellence for Engineering Applications'' (grant reference 823691), the Foundation for Strategic Research (SSF) via the Infrastructure Fellow programme, and EuroCC Project which has received funding from the European Union’s Horizon 2020 research and innovation programme under Grant 951732. JY and FH are funded by TTU Distinguished Chair.
Part of the computations were enabled by resources provided by the Swedish National Infrastructure for Computing (SNIC), partially funded by the Swedish Research Council through grant agreement no. 2018-05973. We also acknowledge computations performed on Longhorn at the Texas Advanced Computing Center (TACC), on Piz Daint at the Swiss National Supercomputing Centre (CSCS), on Berzelius at the National Supercomputer Centre (NSC) and on JUWELS Booster at the J\"ulich Supercomputing Centre (JSC).
\end{acks}

\bibliographystyle{ACM-Reference-Format}
\bibliography{refs}


\begin{thebibliography}{43}


\ifx \showCODEN    \undefined \def \showCODEN     #1{\unskip}     \fi
\ifx \showDOI      \undefined \def \showDOI       #1{#1}\fi
\ifx \showISBNx    \undefined \def \showISBNx     #1{\unskip}     \fi
\ifx \showISBNxiii \undefined \def \showISBNxiii  #1{\unskip}     \fi
\ifx \showISSN     \undefined \def \showISSN      #1{\unskip}     \fi
\ifx \showLCCN     \undefined \def \showLCCN      #1{\unskip}     \fi
\ifx \shownote     \undefined \def \shownote      #1{#1}          \fi
\ifx \showarticletitle \undefined \def \showarticletitle #1{#1}   \fi
\ifx \showURL      \undefined \def \showURL       {\relax}        \fi
\providecommand\bibfield[2]{#2}
\providecommand\bibinfo[2]{#2}
\providecommand\natexlab[1]{#1}
\providecommand\showeprint[2][]{arXiv:#2}

\bibitem[\protect\citeauthoryear{Amdahl}{Amdahl}{1967}]%
        {amdahl1967validity}
\bibfield{author}{\bibinfo{person}{Gene~M Amdahl}.}
  \bibinfo{year}{1967}\natexlab{}.
\newblock \showarticletitle{Validity of the single processor approach to
  achieving large scale computing capabilities}. In
  \bibinfo{booktitle}{\emph{Proceedings of the April 18-20, 1967, spring joint
  computer conference}}. \bibinfo{pages}{483--485}.
\newblock


\bibitem[\protect\citeauthoryear{Choquette, Giroux, and Foley}{Choquette
  et~al\mbox{.}}{2018}]%
        {choquette2018volta}
\bibfield{author}{\bibinfo{person}{Jack Choquette}, \bibinfo{person}{Olivier
  Giroux}, {and} \bibinfo{person}{Denis Foley}.}
  \bibinfo{year}{2018}\natexlab{}.
\newblock \showarticletitle{Volta: Performance and programmability}.
\newblock \bibinfo{journal}{\emph{Ieee Micro}} \bibinfo{volume}{38},
  \bibinfo{number}{2} (\bibinfo{year}{2018}), \bibinfo{pages}{42--52}.
\newblock


\bibitem[\protect\citeauthoryear{Deakin, McIntosh-Smith, Price, Poenaru,
  Atkinson, Popa, and Salmon}{Deakin et~al\mbox{.}}{2019}]%
        {deakin2019performance}
\bibfield{author}{\bibinfo{person}{Tom Deakin}, \bibinfo{person}{Simon
  McIntosh-Smith}, \bibinfo{person}{James Price}, \bibinfo{person}{Andrei
  Poenaru}, \bibinfo{person}{Patrick Atkinson}, \bibinfo{person}{Codrin Popa},
  {and} \bibinfo{person}{Justin Salmon}.} \bibinfo{year}{2019}\natexlab{}.
\newblock \showarticletitle{Performance portability across diverse computer
  architectures}. In \bibinfo{booktitle}{\emph{2019 IEEE/ACM International
  Workshop on Performance, Portability and Productivity in HPC (P3HPC)}}. IEEE,
  \bibinfo{pages}{1--13}.
\newblock


\bibitem[\protect\citeauthoryear{Dennard, Gaensslen, Yu, Rideout, Bassous, and
  LeBlanc}{Dennard et~al\mbox{.}}{1974}]%
        {dennard1974design}
\bibfield{author}{\bibinfo{person}{Robert~H Dennard}, \bibinfo{person}{Fritz~H
  Gaensslen}, \bibinfo{person}{Hwa-Nien Yu}, \bibinfo{person}{V~Leo Rideout},
  \bibinfo{person}{Ernest Bassous}, {and} \bibinfo{person}{Andre~R LeBlanc}.}
  \bibinfo{year}{1974}\natexlab{}.
\newblock \showarticletitle{Design of ion-implanted MOSFET's with very small
  physical dimensions}.
\newblock \bibinfo{journal}{\emph{IEEE Journal of Solid-State Circuits}}
  \bibinfo{volume}{9}, \bibinfo{number}{5} (\bibinfo{year}{1974}),
  \bibinfo{pages}{256--268}.
\newblock


\bibitem[\protect\citeauthoryear{Dongarra, Beckman, Moore, Aerts, Aloisio,
  Andre, Barkai, Berthou, Boku, Braunschweig, et~al\mbox{.}}{Dongarra
  et~al\mbox{.}}{2011}]%
        {dongarra2011international}
\bibfield{author}{\bibinfo{person}{Jack Dongarra}, \bibinfo{person}{Pete
  Beckman}, \bibinfo{person}{Terry Moore}, \bibinfo{person}{Patrick Aerts},
  \bibinfo{person}{Giovanni Aloisio}, \bibinfo{person}{Jean-Claude Andre},
  \bibinfo{person}{David Barkai}, \bibinfo{person}{Jean-Yves Berthou},
  \bibinfo{person}{Taisuke Boku}, \bibinfo{person}{Bertrand Braunschweig},
  {et~al\mbox{.}}} \bibinfo{year}{2011}\natexlab{}.
\newblock \showarticletitle{The international exascale software project
  roadmap}.
\newblock \bibinfo{journal}{\emph{The international journal of high performance
  computing applications}} \bibinfo{volume}{25}, \bibinfo{number}{1}
  (\bibinfo{year}{2011}), \bibinfo{pages}{3--60}.
\newblock


\bibitem[\protect\citeauthoryear{Eichst{\"a}dt, Vymazal, Moxey, and
  Peir{\'o}}{Eichst{\"a}dt et~al\mbox{.}}{2020}]%
        {eichstadt2020comparison}
\bibfield{author}{\bibinfo{person}{Jan Eichst{\"a}dt}, \bibinfo{person}{Martin
  Vymazal}, \bibinfo{person}{David Moxey}, {and} \bibinfo{person}{Joaquim
  Peir{\'o}}.} \bibinfo{year}{2020}\natexlab{}.
\newblock \showarticletitle{A comparison of the shared-memory parallel
  programming models OpenMP, OpenACC and Kokkos in the context of implicit
  solvers for high-order FEM}.
\newblock \bibinfo{journal}{\emph{Computer Physics Communications}}
  \bibinfo{volume}{255} (\bibinfo{year}{2020}), \bibinfo{pages}{107245}.
\newblock


\bibitem[\protect\citeauthoryear{El~Khoury, Schlatter, Noorani, Fischer,
  Brethouwer, and Johansson}{El~Khoury et~al\mbox{.}}{2013}]%
        {elkhoury_schlatter_noorani_fischer_brethouwer_johansson_2013}
\bibfield{author}{\bibinfo{person}{G.~K. El~Khoury}, \bibinfo{person}{P.
  Schlatter}, \bibinfo{person}{A. Noorani}, \bibinfo{person}{P.~F. Fischer},
  \bibinfo{person}{G. Brethouwer}, {and} \bibinfo{person}{A.~V. Johansson}.}
  \bibinfo{year}{2013}\natexlab{}.
\newblock \showarticletitle{Direct numerical simulation of turbulent pipe flows
  at moderately high {R}eynolds numbers}.
\newblock \bibinfo{journal}{\emph{Flow Turbulence Combust.}}
  \bibinfo{volume}{91} (\bibinfo{year}{2013}), \bibinfo{pages}{475--495}.
\newblock


\bibitem[\protect\citeauthoryear{{Fischer}}{{Fischer}}{1998}]%
        {Fischer1998}
\bibfield{author}{\bibinfo{person}{P. {Fischer}}.}
  \bibinfo{year}{1998}\natexlab{}.
\newblock \showarticletitle{{Projection techniques for iterative solution of $A
  \underline{x}=\underline{b}$ with successive right-hand sides}}.
\newblock \bibinfo{journal}{\emph{Computer Methods in Applied Mechanics and
  Engineering}}  \bibinfo{volume}{163} (\bibinfo{date}{Sept.}
  \bibinfo{year}{1998}), \bibinfo{pages}{193--204}.
\newblock
\urldef\tempurl%
\url{https://doi.org/10.1016/S0045-7825(98)00012-7}
\showDOI{\tempurl}


\bibitem[\protect\citeauthoryear{Fischer, Kerkemeier, Min, Lan, Phillips,
  Rathnayake, Merzari, Tomboulides, Karakus, Chalmers, et~al\mbox{.}}{Fischer
  et~al\mbox{.}}{2021}]%
        {fischer2021nekrs}
\bibfield{author}{\bibinfo{person}{Paul Fischer}, \bibinfo{person}{Stefan
  Kerkemeier}, \bibinfo{person}{Misun Min}, \bibinfo{person}{Yu-Hsiang Lan},
  \bibinfo{person}{Malachi Phillips}, \bibinfo{person}{Thilina Rathnayake},
  \bibinfo{person}{Elia Merzari}, \bibinfo{person}{Ananias Tomboulides},
  \bibinfo{person}{Ali Karakus}, \bibinfo{person}{Noel Chalmers},
  {et~al\mbox{.}}} \bibinfo{year}{2021}\natexlab{}.
\newblock \showarticletitle{NekRS, a GPU-Accelerated Spectral Element
  Navier-Stokes Solver}.
\newblock \bibinfo{journal}{\emph{arXiv preprint arXiv:2104.05829}}
  (\bibinfo{year}{2021}).
\newblock


\bibitem[\protect\citeauthoryear{Fischer, Min, Rathnayake, Dutta, Kolev,
  Dobrev, Camier, Kronbichler, Warburton, {\'S}wirydowicz,
  et~al\mbox{.}}{Fischer et~al\mbox{.}}{2020}]%
        {fischer2020scalability}
\bibfield{author}{\bibinfo{person}{Paul Fischer}, \bibinfo{person}{Misun Min},
  \bibinfo{person}{Thilina Rathnayake}, \bibinfo{person}{Som Dutta},
  \bibinfo{person}{Tzanio Kolev}, \bibinfo{person}{Veselin Dobrev},
  \bibinfo{person}{Jean-Sylvain Camier}, \bibinfo{person}{Martin Kronbichler},
  \bibinfo{person}{Tim Warburton}, \bibinfo{person}{Kasia {\'S}wirydowicz},
  {et~al\mbox{.}}} \bibinfo{year}{2020}\natexlab{}.
\newblock \showarticletitle{Scalability of high-performance PDE solvers}.
\newblock \bibinfo{journal}{\emph{The International Journal of High Performance
  Computing Applications}} \bibinfo{volume}{34}, \bibinfo{number}{5}
  (\bibinfo{year}{2020}), \bibinfo{pages}{562--586}.
\newblock


\bibitem[\protect\citeauthoryear{Fischer, Heisey, and Min}{Fischer
  et~al\mbox{.}}{2015}]%
        {fischer2015scaling}
\bibfield{author}{\bibinfo{person}{P.~F. Fischer}, \bibinfo{person}{K. Heisey},
  {and} \bibinfo{person}{M. Min}.} \bibinfo{year}{2015}\natexlab{}.
\newblock \showarticletitle{Scaling Limits for PDE-Based Simulation (Invited)}.
  In \bibinfo{booktitle}{\emph{AIAA Aviation. American Institute of Aeronautics
  and Astronautics}}. \bibinfo{pages}{AIAA 2015--3049}.
\newblock


\bibitem[\protect\citeauthoryear{Fischer, Lottes, and Kerkemeier}{Fischer
  et~al\mbox{.}}{2008}]%
        {fischer2008nek5000}
\bibfield{author}{\bibinfo{person}{Paul~F Fischer}, \bibinfo{person}{James~W
  Lottes}, {and} \bibinfo{person}{Stefan~G Kerkemeier}.}
  \bibinfo{year}{2008}\natexlab{}.
\newblock \bibinfo{title}{nek5000 Web page}.
\newblock
\newblock


\bibitem[\protect\citeauthoryear{Foley and Danskin}{Foley and Danskin}{2017}]%
        {foley2017ultra}
\bibfield{author}{\bibinfo{person}{Denis Foley} {and} \bibinfo{person}{John
  Danskin}.} \bibinfo{year}{2017}\natexlab{}.
\newblock \showarticletitle{Ultra-performance Pascal GPU and NVLink
  interconnect}.
\newblock \bibinfo{journal}{\emph{IEEE Micro}} \bibinfo{volume}{37},
  \bibinfo{number}{2} (\bibinfo{year}{2017}), \bibinfo{pages}{7--17}.
\newblock


\bibitem[\protect\citeauthoryear{Gong, Markidis, Laure, Otten, Fischer, and
  Min}{Gong et~al\mbox{.}}{2016}]%
        {gong2016nekbone}
\bibfield{author}{\bibinfo{person}{Jing Gong}, \bibinfo{person}{Stefano
  Markidis}, \bibinfo{person}{Erwin Laure}, \bibinfo{person}{Matthew Otten},
  \bibinfo{person}{Paul Fischer}, {and} \bibinfo{person}{Misun Min}.}
  \bibinfo{year}{2016}\natexlab{}.
\newblock \showarticletitle{Nekbone performance on {GPU}s with {OpenACC} and
  {CUDA} {F}ortran implementations}.
\newblock \bibinfo{journal}{\emph{The Journal of Supercomputing}}
  \bibinfo{volume}{72}, \bibinfo{number}{11} (\bibinfo{year}{2016}),
  \bibinfo{pages}{4160--4180}.
\newblock


\bibitem[\protect\citeauthoryear{Hariri, Tran, Jocksch, Lanti, Progsch,
  Messmer, Brunner, Gheller, and Villard}{Hariri et~al\mbox{.}}{2016}]%
        {hariri2016portable}
\bibfield{author}{\bibinfo{person}{Farah Hariri}, \bibinfo{person}{Trach-Minh
  Tran}, \bibinfo{person}{Andreas Jocksch}, \bibinfo{person}{Emmanuel Lanti},
  \bibinfo{person}{J Progsch}, \bibinfo{person}{Peter Messmer},
  \bibinfo{person}{Stephan Brunner}, \bibinfo{person}{Claudio Gheller}, {and}
  \bibinfo{person}{Laurent Villard}.} \bibinfo{year}{2016}\natexlab{}.
\newblock \showarticletitle{A portable platform for accelerated {PIC} codes and
  its application to {GPU}s using {OpenACC}}.
\newblock \bibinfo{journal}{\emph{Computer Physics Communications}}
  \bibinfo{volume}{207} (\bibinfo{year}{2016}), \bibinfo{pages}{69--82}.
\newblock
\urldef\tempurl%
\url{https://doi.org/10.1016/j.cpc.2016.05.008]}
\showDOI{\tempurl}


\bibitem[\protect\citeauthoryear{Jansson, Karp, Podobas, Markidis, and
  Schlatter}{Jansson et~al\mbox{.}}{2021}]%
        {jansson2021neko}
\bibfield{author}{\bibinfo{person}{Niclas Jansson}, \bibinfo{person}{Martin
  Karp}, \bibinfo{person}{Artur Podobas}, \bibinfo{person}{Stefano Markidis},
  {and} \bibinfo{person}{Philipp Schlatter}.} \bibinfo{year}{2021}\natexlab{}.
\newblock \showarticletitle{Neko: A Modern, Portable, and Scalable Framework
  for High-Fidelity Computational Fluid Dynamics}.
\newblock \bibinfo{journal}{\emph{arXiv preprint arXiv:2107.01243}}
  (\bibinfo{year}{2021}).
\newblock


\bibitem[\protect\citeauthoryear{Jocksch, Kraushaar, and Daverio}{Jocksch
  et~al\mbox{.}}{2019}]%
        {jocksch2019optimized}
\bibfield{author}{\bibinfo{person}{Andreas Jocksch}, \bibinfo{person}{Matthias
  Kraushaar}, {and} \bibinfo{person}{David Daverio}.}
  \bibinfo{year}{2019}\natexlab{}.
\newblock \showarticletitle{Optimized all-to-all communication on multicore
  architectures applied to FFTs with pencil decomposition}.
\newblock \bibinfo{journal}{\emph{Concurrency and Computation: Practice and
  Experience}} \bibinfo{volume}{31}, \bibinfo{number}{16}
  (\bibinfo{year}{2019}), \bibinfo{pages}{e4964}.
\newblock


\bibitem[\protect\citeauthoryear{Karp, Jansson, Podobas, Schlatter, and
  Markidis}{Karp et~al\mbox{.}}{2020}]%
        {karp2020optimization}
\bibfield{author}{\bibinfo{person}{Martin Karp}, \bibinfo{person}{Niclas
  Jansson}, \bibinfo{person}{Artur Podobas}, \bibinfo{person}{Philipp
  Schlatter}, {and} \bibinfo{person}{Stefano Markidis}.}
  \bibinfo{year}{2020}\natexlab{}.
\newblock \showarticletitle{Optimization of tensor-product operations in
  nekbone on gpus}.
\newblock \bibinfo{journal}{\emph{arXiv preprint arXiv:2005.13425}}
  (\bibinfo{year}{2020}).
\newblock


\bibitem[\protect\citeauthoryear{Karp, Podobas, Jansson, Kenter, Plessl,
  Schlatter, and Markidis}{Karp et~al\mbox{.}}{2021}]%
        {karp2021High}
\bibfield{author}{\bibinfo{person}{Martin Karp}, \bibinfo{person}{Artur
  Podobas}, \bibinfo{person}{Niclas Jansson}, \bibinfo{person}{Tobias Kenter},
  \bibinfo{person}{Christian Plessl}, \bibinfo{person}{Philipp Schlatter},
  {and} \bibinfo{person}{Stefano Markidis}.} \bibinfo{year}{2021}\natexlab{}.
\newblock \showarticletitle{High-Performance Spectral Element Methods on
  Field-Programmable Gate Arrays : Implementation, Evaluation, and Future
  Projection}. In \bibinfo{booktitle}{\emph{2021 IEEE International Parallel
  and Distributed Processing Symposium (IPDPS)}}. \bibinfo{pages}{1077--1086}.
\newblock
\urldef\tempurl%
\url{https://doi.org/10.1109/IPDPS49936.2021.00116}
\showDOI{\tempurl}


\bibitem[\protect\citeauthoryear{Lee and Moser}{Lee and Moser}{2015}]%
        {myoungkyu2015JFM}
\bibfield{author}{\bibinfo{person}{Myoungkyu Lee} {and}
  \bibinfo{person}{Robert~D Moser}.} \bibinfo{year}{2015}\natexlab{}.
\newblock \showarticletitle{Direct numerical simulation of turbulent channel
  flow up to $Re_\tau =5200$}.
\newblock \bibinfo{journal}{\emph{Journal of Fluid Mechanics}}
  \bibinfo{volume}{774} (\bibinfo{year}{2015}), \bibinfo{pages}{395--415}.
\newblock


\bibitem[\protect\citeauthoryear{Markidis, Gong, Schliephake, Laure, Hart,
  Henty, Heisey, and Fischer}{Markidis et~al\mbox{.}}{2015}]%
        {markidis2015openacc}
\bibfield{author}{\bibinfo{person}{Stefano Markidis}, \bibinfo{person}{Jing
  Gong}, \bibinfo{person}{Michael Schliephake}, \bibinfo{person}{Erwin Laure},
  \bibinfo{person}{Alistair Hart}, \bibinfo{person}{David Henty},
  \bibinfo{person}{Katherine Heisey}, {and} \bibinfo{person}{Paul Fischer}.}
  \bibinfo{year}{2015}\natexlab{}.
\newblock \showarticletitle{OpenACC acceleration of the Nek5000 spectral
  element code}.
\newblock \bibinfo{journal}{\emph{The International Journal of High Performance
  Computing Applications}} \bibinfo{volume}{29}, \bibinfo{number}{3}
  (\bibinfo{year}{2015}), \bibinfo{pages}{311--319}.
\newblock


\bibitem[\protect\citeauthoryear{Medina, St-Cyr, and Warburton}{Medina
  et~al\mbox{.}}{2014}]%
        {medina2014occa}
\bibfield{author}{\bibinfo{person}{David~S Medina}, \bibinfo{person}{Amik
  St-Cyr}, {and} \bibinfo{person}{Tim Warburton}.}
  \bibinfo{year}{2014}\natexlab{}.
\newblock \showarticletitle{OCCA: A unified approach to multi-threading
  languages}.
\newblock \bibinfo{journal}{\emph{arXiv preprint arXiv:1403.0968}}
  (\bibinfo{year}{2014}).
\newblock


\bibitem[\protect\citeauthoryear{Merzari, Obabko, Fischer, Halford, Walker,
  Siegel, and Yu}{Merzari et~al\mbox{.}}{2017}]%
        {merzari2017large}
\bibfield{author}{\bibinfo{person}{Elia Merzari}, \bibinfo{person}{Aleks
  Obabko}, \bibinfo{person}{Paul Fischer}, \bibinfo{person}{Noah Halford},
  \bibinfo{person}{Justin Walker}, \bibinfo{person}{Andrew Siegel}, {and}
  \bibinfo{person}{Yiqi Yu}.} \bibinfo{year}{2017}\natexlab{}.
\newblock \showarticletitle{Large-scale large eddy simulation of nuclear
  reactor flows: Issues and perspectives}.
\newblock \bibinfo{journal}{\emph{Nuclear Engineering and Design}}
  \bibinfo{volume}{312} (\bibinfo{year}{2017}), \bibinfo{pages}{86--98}.
\newblock
\urldef\tempurl%
\url{https://doi.org/10.1016/j.nucengdes.2016.09.028}
\showDOI{\tempurl}


\bibitem[\protect\citeauthoryear{Munshi}{Munshi}{2009}]%
        {munshi2009opencl}
\bibfield{author}{\bibinfo{person}{Aaftab Munshi}.}
  \bibinfo{year}{2009}\natexlab{}.
\newblock \showarticletitle{The opencl specification}. In
  \bibinfo{booktitle}{\emph{2009 IEEE Hot Chips 21 Symposium (HCS)}}. IEEE,
  \bibinfo{pages}{1--314}.
\newblock


\bibitem[\protect\citeauthoryear{Nvidia}{Nvidia}{2021}]%
        {AMGXweb}
\bibfield{author}{\bibinfo{person}{Nvidia}.} \bibinfo{year}{2021}\natexlab{}.
\newblock \bibinfo{title}{AMGX website}.
\newblock \bibinfo{howpublished}{\url{https://developer.nvidia.com/amgx}}.
\newblock


\bibitem[\protect\citeauthoryear{Nvidia}{Nvidia}{2007}]%
        {Nvidia2007compute}
\bibfield{author}{\bibinfo{person}{CUDA Nvidia}.}
  \bibinfo{year}{2007}\natexlab{}.
\newblock \showarticletitle{Compute unified device architecture programming
  guide}.
\newblock  (\bibinfo{year}{2007}).
\newblock


\bibitem[\protect\citeauthoryear{Offermans, Marin, Schanen, Gong, Fischer,
  Schlatter, Obabko, Peplinski, Hutchinson, and Merzari}{Offermans
  et~al\mbox{.}}{2016}]%
        {offermans2016strong}
\bibfield{author}{\bibinfo{person}{Nicolas Offermans}, \bibinfo{person}{Oana
  Marin}, \bibinfo{person}{Michel Schanen}, \bibinfo{person}{Jing Gong},
  \bibinfo{person}{Paul Fischer}, \bibinfo{person}{Philipp Schlatter},
  \bibinfo{person}{Aleks Obabko}, \bibinfo{person}{Adam Peplinski},
  \bibinfo{person}{Maxwell Hutchinson}, {and} \bibinfo{person}{Elia Merzari}.}
  \bibinfo{year}{2016}\natexlab{}.
\newblock \showarticletitle{On the strong scaling of the spectral element
  solver Nek5000 on petascale systems}. In
  \bibinfo{booktitle}{\emph{Proceedings of the Exascale Applications and
  Software Conference 2016}}. \bibinfo{pages}{1--10}.
\newblock


\bibitem[\protect\citeauthoryear{Otero, Gong, Min, Fischer, Schlatter, and
  Laure}{Otero et~al\mbox{.}}{2019}]%
        {otero_gong_min_fischer_schlatter_laure_2019}
\bibfield{author}{\bibinfo{person}{E. Otero}, \bibinfo{person}{J. Gong},
  \bibinfo{person}{M. Min}, \bibinfo{person}{P.~F. Fischer},
  \bibinfo{person}{P. Schlatter}, {and} \bibinfo{person}{E. Laure}.}
  \bibinfo{year}{2019}\natexlab{}.
\newblock \showarticletitle{{OpenACC} acceleration for the {$P_N-P_{N-2}$}
  algorithm in {Nek5000}}.
\newblock \bibinfo{journal}{\emph{J. Parallel Dist. Comput.}}
  \bibinfo{volume}{132} (\bibinfo{year}{2019}), \bibinfo{pages}{69--78}.
\newblock


\bibitem[\protect\citeauthoryear{{\"O}zg{\"o}kmen, Fischer, and
  Johns}{{\"O}zg{\"o}kmen et~al\mbox{.}}{2006}]%
        {ozgokmen2006product}
\bibfield{author}{\bibinfo{person}{Tamay~M {\"O}zg{\"o}kmen},
  \bibinfo{person}{Paul~F Fischer}, {and} \bibinfo{person}{William~E Johns}.}
  \bibinfo{year}{2006}\natexlab{}.
\newblock \showarticletitle{Product water mass formation by turbulent density
  currents from a high-order nonhydrostatic spectral element model}.
\newblock \bibinfo{journal}{\emph{Ocean Modelling}} \bibinfo{volume}{12},
  \bibinfo{number}{3-4} (\bibinfo{year}{2006}), \bibinfo{pages}{237--267}.
\newblock


\bibitem[\protect\citeauthoryear{Patera}{Patera}{1984}]%
        {patera1984spectral}
\bibfield{author}{\bibinfo{person}{Anthony~T Patera}.}
  \bibinfo{year}{1984}\natexlab{}.
\newblock \showarticletitle{A spectral element method for fluid dynamics:
  laminar flow in a channel expansion}.
\newblock \bibinfo{journal}{\emph{Journal of computational Physics}}
  \bibinfo{volume}{54}, \bibinfo{number}{3} (\bibinfo{year}{1984}),
  \bibinfo{pages}{468--488}.
\newblock


\bibitem[\protect\citeauthoryear{Perot}{Perot}{1993}]%
        {Perot1993}
\bibfield{author}{\bibinfo{person}{J.Blair Perot}.}
  \bibinfo{year}{1993}\natexlab{}.
\newblock \showarticletitle{An Analysis of the Fractional Step Method}.
\newblock \bibinfo{journal}{\emph{J. Comput. Phys.}} \bibinfo{volume}{108},
  \bibinfo{number}{1} (\bibinfo{year}{1993}), \bibinfo{pages}{51 -- 58}.
\newblock
\showISSN{0021-9991}
\urldef\tempurl%
\url{https://doi.org/10.1006/jcph.1993.1162}
\showDOI{\tempurl}


\bibitem[\protect\citeauthoryear{Podobas, Sano, and Matsuoka}{Podobas
  et~al\mbox{.}}{2020}]%
        {podobas2020survey}
\bibfield{author}{\bibinfo{person}{Artur Podobas}, \bibinfo{person}{Kentaro
  Sano}, {and} \bibinfo{person}{Satoshi Matsuoka}.}
  \bibinfo{year}{2020}\natexlab{}.
\newblock \showarticletitle{A survey on coarse-grained reconfigurable
  architectures from a performance perspective}.
\newblock \bibinfo{journal}{\emph{IEEE Access}}  \bibinfo{volume}{8}
  (\bibinfo{year}{2020}), \bibinfo{pages}{146719--146743}.
\newblock


\bibitem[\protect\citeauthoryear{R, Kolvev, Li, Osborn, Osei-Kuffuor, Magri,
  Schroder, Sjogreen, Vassilevski, and Yang}{R et~al\mbox{.}}{2021}]%
        {HYPREweb}
\bibfield{author}{\bibinfo{person}{Fakgout R}, \bibinfo{person}{T. Kolvev},
  \bibinfo{person}{R. Li}, \bibinfo{person}{S. Osborn}, \bibinfo{person}{D.
  Osei-Kuffuor}, \bibinfo{person}{V.~P. Magri}, \bibinfo{person}{J. Schroder},
  \bibinfo{person}{B. Sjogreen}, \bibinfo{person}{P. Vassilevski}, {and}
  \bibinfo{person}{U.~M. Yang}.} \bibinfo{year}{2021}\natexlab{}.
\newblock \bibinfo{title}{HYPRE Website}.
\newblock
  \bibinfo{howpublished}{\url{https://computing.llnl.gov/projects/hypre-scalable-linear-solvers-multigrid-methods}}.
\newblock


\bibitem[\protect\citeauthoryear{Rezaeiravesh, Vinuesa, and
  Schlatter1}{Rezaeiravesh et~al\mbox{.}}{2019}]%
        {saleh2019KTH}
\bibfield{author}{\bibinfo{person}{Saleh Rezaeiravesh},
  \bibinfo{person}{Ricardo Vinuesa}, {and} \bibinfo{person}{Philipp
  Schlatter1}.} \bibinfo{year}{2019}\natexlab{}.
\newblock \bibinfo{booktitle}{\emph{A statistics toolbox for turbulent pipe
  flow in Nek5000}}.
\newblock \bibinfo{type}{{T}echnical {R}eport}. \bibinfo{institution}{KTH
  Technical Report, TRITA-SCI-RAP 2019:008}.
\newblock


\bibitem[\protect\citeauthoryear{Schlatter and Khoury}{Schlatter and
  Khoury}{2012}]%
        {schlatter2012PDC}
\bibfield{author}{\bibinfo{person}{Philipp Schlatter} {and}
  \bibinfo{person}{George K.~El Khoury}.} \bibinfo{year}{2012}\natexlab{}.
\newblock \showarticletitle{Turbulent flow in pipes}.
\newblock \bibinfo{journal}{\emph{PDC newsletter}} (\bibinfo{year}{2012}),
  \bibinfo{pages}{3--10}.
\newblock


\bibitem[\protect\citeauthoryear{Shpiner, Haramaty, Eliad, Zdornov, Gafni, and
  Zahavi}{Shpiner et~al\mbox{.}}{2017}]%
        {shpiner2017dragonfly}
\bibfield{author}{\bibinfo{person}{Alexander Shpiner}, \bibinfo{person}{Zachy
  Haramaty}, \bibinfo{person}{Saar Eliad}, \bibinfo{person}{Vladimir Zdornov},
  \bibinfo{person}{Barak Gafni}, {and} \bibinfo{person}{Eitan Zahavi}.}
  \bibinfo{year}{2017}\natexlab{}.
\newblock \showarticletitle{Dragonfly+: Low cost topology for scaling
  datacenters}. In \bibinfo{booktitle}{\emph{2017 IEEE 3rd International
  Workshop on High-Performance Interconnection Networks in the Exascale and
  Big-Data Era (HiPINEB)}}. IEEE, \bibinfo{pages}{1--8}.
\newblock


\bibitem[\protect\citeauthoryear{{\'S}wirydowicz, Chalmers, Karakus, and
  Warburton}{{\'S}wirydowicz et~al\mbox{.}}{2019}]%
        {swirydowicz2019acceleration}
\bibfield{author}{\bibinfo{person}{Kasia {\'S}wirydowicz},
  \bibinfo{person}{Noel Chalmers}, \bibinfo{person}{Ali Karakus}, {and}
  \bibinfo{person}{Tim Warburton}.} \bibinfo{year}{2019}\natexlab{}.
\newblock \showarticletitle{Acceleration of tensor-product operations for
  high-order finite element methods}.
\newblock \bibinfo{journal}{\emph{The International Journal of High Performance
  Computing Applications}} \bibinfo{volume}{33}, \bibinfo{number}{4}
  (\bibinfo{year}{2019}), \bibinfo{pages}{735--757}.
\newblock


\bibitem[\protect\citeauthoryear{Theis and Wong}{Theis and Wong}{2017}]%
        {theis2017end}
\bibfield{author}{\bibinfo{person}{Thomas~N Theis} {and}
  \bibinfo{person}{H-S~Philip Wong}.} \bibinfo{year}{2017}\natexlab{}.
\newblock \showarticletitle{The end of moore's law: A new beginning for
  information technology}.
\newblock \bibinfo{journal}{\emph{Computing in Science \& Engineering}}
  \bibinfo{volume}{19}, \bibinfo{number}{2} (\bibinfo{year}{2017}),
  \bibinfo{pages}{41--50}.
\newblock
\urldef\tempurl%
\url{https://doi.org/10.1109/MCSE.2017.29}
\showDOI{\tempurl}


\bibitem[\protect\citeauthoryear{Tufo and Fischer}{Tufo and Fischer}{2001}]%
        {Tufo2001}
\bibfield{author}{\bibinfo{person}{H.M Tufo} {and} \bibinfo{person}{P.F
  Fischer}.} \bibinfo{year}{2001}\natexlab{}.
\newblock \showarticletitle{Fast Parallel Direct Solvers for Coarse Grid
  Problems}.
\newblock \bibinfo{journal}{\emph{J. Parallel Distrib. Comput.}}
  \bibinfo{volume}{61}, \bibinfo{number}{2} (\bibinfo{date}{Feb.}
  \bibinfo{year}{2001}), \bibinfo{pages}{151--177}.
\newblock
\showISSN{0743-7315}
\urldef\tempurl%
\url{https://doi.org/10.1006/jpdc.2000.1676}
\showDOI{\tempurl}


\bibitem[\protect\citeauthoryear{Tufo and Fischer}{Tufo and Fischer}{1999}]%
        {tufo1999terascale}
\bibfield{author}{\bibinfo{person}{Henry~M Tufo} {and} \bibinfo{person}{Paul~F
  Fischer}.} \bibinfo{year}{1999}\natexlab{}.
\newblock \showarticletitle{Terascale spectral element algorithms and
  implementations}. In \bibinfo{booktitle}{\emph{Proceedings of the 1999
  ACM/IEEE Conference on Supercomputing}}. \bibinfo{pages}{68--es}.
\newblock
\urldef\tempurl%
\url{https://doi.org/10.1145/331532.331599}
\showDOI{\tempurl}


\bibitem[\protect\citeauthoryear{Vinuesa, Negi, Atzori, Hanifi, Henningson, and
  Schlatter}{Vinuesa et~al\mbox{.}}{2018}]%
        {vinuesa2018turbulent}
\bibfield{author}{\bibinfo{person}{R. Vinuesa}, \bibinfo{person}{P.S. Negi},
  \bibinfo{person}{M. Atzori}, \bibinfo{person}{A. Hanifi},
  \bibinfo{person}{D.S. Henningson}, {and} \bibinfo{person}{P. Schlatter}.}
  \bibinfo{year}{2018}\natexlab{}.
\newblock \showarticletitle{Turbulent boundary layers around wing sections up
  to Rec=1,000,000}.
\newblock \bibinfo{journal}{\emph{International Journal of Heat and Fluid
  Flow}}  \bibinfo{volume}{72} (\bibinfo{year}{2018}), \bibinfo{pages}{86--99}.
\newblock
\showISSN{0142-727X}
\urldef\tempurl%
\url{https://doi.org/10.1016/j.ijheatfluidflow.2018.04.017}
\showDOI{\tempurl}


\bibitem[\protect\citeauthoryear{Walker and Dongarra}{Walker and
  Dongarra}{1996}]%
        {walker1996mpi}
\bibfield{author}{\bibinfo{person}{David~W Walker} {and}
  \bibinfo{person}{Jack~J Dongarra}.} \bibinfo{year}{1996}\natexlab{}.
\newblock \showarticletitle{MPI: a standard message passing interface}.
\newblock \bibinfo{journal}{\emph{Supercomputer}}  \bibinfo{volume}{12}
  (\bibinfo{year}{1996}), \bibinfo{pages}{56--68}.
\newblock


\bibitem[\protect\citeauthoryear{Willis}{Willis}{2017}]%
        {openpipeflow}
\bibfield{author}{\bibinfo{person}{Ashley~P Willis}.}
  \bibinfo{year}{2017}\natexlab{}.
\newblock \showarticletitle{The {O}penpipeflow {N}avier--{S}tokes Solver}.
\newblock \bibinfo{journal}{\emph{SoftwareX}}  \bibinfo{volume}{6}
  (\bibinfo{year}{2017}), \bibinfo{pages}{124--127}.
\newblock
\urldef\tempurl%
\url{https://doi.org/10.1016/j.softx.2017.05.003}
\showDOI{\tempurl}


\end{thebibliography}

\end{document}